# Sensing Throughput Tradeoff for Cognitive Radio Networks with Noise Variance Uncertainty

Tadilo Endeshaw Bogale[+], Luc Vandendorpe[++] and Long Bao Le[+]
[+]Institute National de la Recherche Scientifique (INRS)
Université du Québec, Montréal, Canada
[++]ICTEAM Institute, Université Catholique de Louvain
Louvain la neuve, Belgium
Email: tadilo.bogale@emt.inrs.ca, luc.vandendorpe@uclouvain.be and long.le@emt.inrs.ca

*Invited paper*

*Abstract*— This paper proposes novel spectrum sensing algorithm, and examines the sensing throughput tradeoff for cognitive radio (CR) networks under noise variance uncertainty. It is assumed that there are one white sub-band, and one target sub-band which is either white or non-white. Under this assumption, first we propose a novel generalized energy detector (GED) for examining the target sub-band by exploiting the noise information of the white sub-band, then, we study the tradeoff between the sensing time and achievable throughput of the CR network. To study this tradeoff, we consider the sensing time optimization for maximizing the throughput of the CR network while appropriately protecting the primary network. The sensing time is optimized by utilizing the derived detection and false alarm probabilities of the GED. The proposed GED does not suffer from signal to noise ratio (SNR) wall (i.e., robust against noise variance uncertainty) and outperforms the existing signal detectors. Moreover, the relationship between the proposed GED and conventional energy detector (CED) is quantified analytically. We show that the optimal sensing times with perfect and imperfect noise variances are not the same. In particular, when the frame duration is 2s, SNR= $-20$dB, and each of the bandwidths of the white and target sub-bands is 6MHz, the optimal sensing times are 28.5ms and 50.6ms with perfect and imperfect noise variances, respectively.

## I. INTRODUCTION

Cognitive radio (CR) is one of the promising approaches to improve the spectral efficiency of current wireless networks [1], [2]. One key feature of a CR network is the potential to learn its surrounding radio environment, which is performed by the spectrum sensing (signal detection) part of the CR device. The most widely known signal detectors are matched filter, energy and cyclostationary based detectors. Among these, the matched filter is optimal, which, however, requires perfect synchronization between the primary transmitter and cognitive device [3]. The energy detector (hereafter referred as conventional energy detector (CED)) does not require any information about the primary user and it is simple to implement. However, the CED is very sensitive to noise variance uncertainty, and there is a signal to noise ratio (SNR) wall below which this detector can not guarantee the desired detection performance [3]–[5]. Cyclostationary based detector is robust against noise variance uncertainty and it can reject the effect of external interference, which unfortunately has high computational complexity and is sensitive to cyclic frequency mismatch [5]–[7].

In [8], the eigenvalue decomposition (EVD)-based signal detector is proposed. This detector is robust against noise variance uncertainty but its computational complexity is high [9], [10]. Recently in [11], new Max-Min SNR based signal detector is proposed. This paper employs linear combination approach of the oversampled received signal. Under noise variance uncertainty, simulation results demonstrate that this detector achieves better performance than those of the CED and EVD-based detectors in additive white Gaussian noise (AWGN) and Rayleigh fading channels. The detector of [11] also guarantee the desired probability of detection (false alarm) $P_d(P_f)$ in the presence of low (moderate) adjacent channel interference (ACI) signals. The main drawbacks of [11], however, are that the cognitive device requires accurate knowledge of the primary transmitter pulse shaping filter and rolloff factor. Furthermore, the approach of [11] employs oversampling of the received signal beyond the Nyquist rate, which may not be desirable in practice as such operation requires expensive high-speed analog to digital converter. In addition, the theoretical $P_f$ and $P_d$ expressions are obtained by employing numerical methods. Nonetheless, as will be clear later, the detection algorithm of the current paper achieves better performance than that of the algorithm of [11] under similar assumptions. Furthermore, the proposed detector of the current paper requires neither the transmitter pulse shaping filter nor oversampling of the received signal.

In [12], throughput maximization problem for CR network has been considered by employing the CED. However, as we have explained previously, the CED suffers from SNR wall. Thus, it may not be possible to utilize the CED to maximize the throughput of the CR network for practically relevant SNR of the transmitted signal (i.e., $-20$dB).

In the current paper we propose novel spectrum sensing algorithm, and examine the sensing throughput tradeoff for CR networks under noise variance uncertainty. It is assumed that there are one white sub-band, and one target sub-band which is either white or non-white. In a wide-band scenario, the white sub-band can be identified by applying the approach of [13] (see [13] for more details). Under this assumption, first we propose a novel generalized energy detector (GED) for examining the target sub-band by exploiting the noise information of the white sub-band, then, we study the tradeoff between the sensing time and achievable throughput of the CR network. To study this tradeoff, we consider the sensing time optimization for maximizing the throughput of the CR network while appropriately protecting the primary network. The sensing time is optimized by utilizing the derived $P_d$ and $P_f$ of the GED. The proposed GED does not suffer from SNR wall (i.e., robust against noise variance uncertainty) and outperforms the existing signal detectors. To optimize the sensing time, we consider that the CR network performs sensing and then transmission periodically over equal frame intervals. This frame-based sensing and transmission strategy has been commonly adopted in the literature [12]. This frame interval can be set equal to the required channel evacuation time which is 2s in the 802.22 standard for example.

The main contributions of the paper are summarized as follows.

- We propose novel generalized energy detection algorithm to detect the target sub-band under noise variance uncertainty. The proposed GED utilizes the noise information collected from the white sub-band. Furthermore, the GED is designed to ensure a certain $P_d(P_f)$ performance without experiencing any SNR wall.
- We derive the $P_d$ and $P_f$ expressions for the proposed GED. These derivations reveal that the detection performance of the GED depends on the bandwidths of the white and target sub-bands. The derivations also exploit the fact that when the noise variance is estimated from finite sub-band, the theoretical thresholds of the CED can not be applied directly. Moreover, the relationship between the proposed GED and CED is quantified analytically.
- We formulate the sensing time optimization problem for the CR network as a concave maximization problem where its solution is obtained by using convex optimization tools. We show that the optimal sensing time obtained by the proposed GED ($T_{oGED}$) is different from that of the CED ($T_{oCED}$). In other words, the optimal sensing durations with perfect and imperfect noise variances are not the same.
- Numerical studies are conducted to investigate the performance of the proposed sensing and optimization framework. Specifically, we validate the analytical results by comparing with computer simulation. We demonstrate that the proposed detection algorithm is robust against noise variance uncertainty and outperforms existing spectrum sensing algorithms. In an exemplifying setting, when SNR= $-20$dB, frame duration is 2s, and the bandwidth of each of the white and target sub-bands is 6MHz, the optimal sensing times that maximizes the throughput are 28.5ms and 50.6ms for perfect and imperfect noise variance scenarios, respectively (i.e., $T_{oGED} = 1.78 T_{oCED}$).

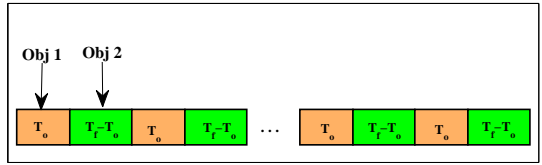

Fig. 1. The frame structure of a cognitive radio network.

The remaining part of this paper is organized as follows: Section II discusses the signal model and problem statement. The proposed generalized energy detection algorithm is discussed in Section III. The sensing time optimization algorithm is presented in Section IV. In Sections V and VI, detailed numerical and simulation results are presented for different practically relevant parameter settings. Finally, conclusions are drawn in Section VII.

*Notations:* The following notations are used: $AE(.)$ denotes the average energy and $\lfloor x \rfloor$ ($\lceil x \rceil$) is the nearest integer less (greater) than or equal to $x$. The representations s.t, $Pr(.)$, $(.)^\star$, $E\{.\}$ and $|.|$ denote subject to, probability, optimal, expectation and absolute value, respectively.

## II. SIGNAL MODEL AND PROBLEM STATEMENT

Consider a cognitive radio network that operates on the spectrum of $B$Hz. This $B$Hz has two sub-bands where one of the sub-band is white and the other sub-band is the target sub-band which could be either white or non-white (i.e., unknown sub-band). We assume that the cognitive device attempts to perform spectrum sensing on the target sub-band and utilizes this sub-band for communications only when it is white[1].

The frame structure of a CR network is illustrated in Fig. 1, which shows consecutive frames. Here, each frame has a duration of $T_f = T_o + (T_f - T_o)$, where $T_o$ is used to sense the target sub-band and the remaining time $T_f - T_o$ is used for CR transmission. The sensing time $T_o$ is required to ensure that the primary network is sufficiently protected. This protection level is usually expressed in terms of $P_d$ of the target sub-band. The time $T_f - T_o$ is usually chosen to ensure that the target sub-band is efficiently exploited. In this paper, we are interested in designing the spectrum sensing to maximize the throughput of the CR network while appropriately protecting the primary network. This problem is formulated as

$$\max Th_f, \quad \text{s.t } P_{df}(\text{target sub band}) \geq \tilde{P}_d \qquad (1)$$

where $Th_f$, $P_{df}(.)$ and $\tilde{P}_d$ are the throughput achieved in the target sub-band, detection probability and required $P_d$ threshold in frame $f$, respectively. As we can see from Fig. 1, when we increase $T_o$, the protection level of the primary network increases (i.e., $P_{df}(.)$ increases). However, doing this will reduce the achievable throughput of the CR network which is directly related to $T_f - T_o$. Therefore, the tradeoff between sensing time and throughput can be studied by examining (1). Furthermore, for the appropriately selected $T_f$, the optimal solution of (1) will satisfy $0 < T_o \leq T_f$. We assume that $T_f$ is selected appropriately and the sampling frequency of the cognitive device is set to $B$ (i.e., Nyquist sampling). Given these assumptions, problem (1) can be solved by addressing the following two objectives for each frame.

---

[1] A CR network is a network that does not have exclusive right to use this sub-band. It is always termed as a secondary network.

**Obj 1**: Detecting the target sub-band using the noise information of the white sub-band by the proposed GED[2].

**Obj 2**: Optimizing the sensing time to maximize the throughput of the CR network by employing the $P_d$ and $P_f$ expressions of the GED.

## III. GENERALIZED ENERGY DETECTOR

This section presents the proposed GED, and provides detailed performance analysis. As mentioned in the above section, we have two sub-bands where one of them is the white sub-band and the other is the target sub-band. For convenience, let us represent the *target sub-band as sub-band $k$* with bandwidth $B_k$ (i.e., from 0 to $B_k Hz$) and the *white sub-band as sub-band $i$* with bandwidth $B_i$ (i.e., from $B_k$ to $BHz$), and we assume that the cognitive device employs $T_{ts}$ sensing time to ensure the target $\tilde{P}_d$.

The base-band received signal of each frame $r(t)$ can thus be expressed as

$$r(t) = s(t) + w(t), \quad 0 \leq t \leq T_{ts} \quad (2)$$

where $s(t)$ and $w(t)$ are the signal and noise components, respectively. By sampling this signal with period $\frac{1}{B}$ (i.e., Nyquist sampling rate), the sampled version of $r(t)$, with slight abuse of notation, can be expressed as

$$r[n] = s[n] + w[n], \quad n = 1, \cdots, N_s \quad (3)$$

where $N_s = T_{ts} B$ is the number of received samples in $T_{ts}$ duration. It is assumed that $w[n], \forall n$ are independent and identically distributed (i.i.d) zero mean circularly symmetric complex Gaussian (ZMCSCG) random variables all with variance $\sigma^2$ which is unknown or known imperfectly.

The discrete time Fourier transform (DFT) of $r[n]$ is given as [14]

$$\tilde{r}[m] = \sum_{n=1}^{N_s} \frac{r[n] \exp \frac{-i 2\pi (m-1)(n-1)}{N_s}}{\sqrt{N_s}}, \quad m = 1, \cdots, N_s.$$

The average energies of the frequency regions $[0 : B_k]$ and $[B_k : B]$ are given by

$$\text{AE}([0 : B_k]) = \sum_{m=1}^{N_{dk}} \frac{|\tilde{r}[m]|^2}{N_{dk}} \triangleq \sum_{j=1}^{N_{dk}} \frac{|d_k[j]|^2}{N_{dk}}$$
$$\triangleq \hat{M}_{dk}$$
$$\text{AE}([B_k : B]) = \sum_{m=1}^{N_z} \frac{|\tilde{r}[N_k + m]|^2}{N_z} \triangleq \sum_{j=1}^{N_z} \frac{|z[j]|^2}{N_z}$$
$$\triangleq \hat{M}_z \quad (4)$$

[2] As will be detailed later, our GED is not a straightforward extension of the CED.

where $N_{dk} = \lfloor \frac{B_k}{B} N_s \rfloor$, $N_z = N_s - N_{dk}$, $\{d_k[j] = \tilde{r}[m]\}_{j=m=1}^{N_{dk}}$ and $\{z[j] = \tilde{r}[N_k + m]\}_{j=m=1}^{N_z}$.

As the $i$th white sub-band contains noise only signal, $z[j], \forall j$ can be modeled as i.i.d ZMCSCG random variables all with variance $\sigma^2$ whereas, the $k$th sub-band contains either noise only ($H_{0k}$) or signal plus noise ($H_{1k}$). Thus, $d_k[j], k \neq i$ can be modeled as

$$d_k[j] = \begin{cases} \tilde{s}_k[j] + \tilde{w}_k[j], & \text{Under } H_{1k} \\ \tilde{w}_k[j], & j = 1, \cdots, N_{dk} \text{ Under } H_{0k} \end{cases} \quad (5)$$

where $\tilde{w}_k[j], \forall j$ are i.i.d ZMCSCG random variables all with variance $\sigma^2$, $\tilde{s}_k[j], \forall j$ are i.i.d zero mean random variables with $\text{E}\{|\tilde{s}_k[j]|^2\} = \gamma_k \sigma^2$ and $\gamma_k$ denotes the SNR of the $k$th sub-band under $H_{1k}$ hypothesis.

To detect the $k$th sub-band, we propose the following test statistics

$$R_k = \sqrt{\frac{N_{dk} \beta_k}{\beta_k + 1}} \left( \frac{\hat{M}_{dk}}{\hat{M}_z} - 1 \right), \quad k \neq i \quad (6)$$

where $\beta_k = \frac{N_z}{N_{dk}} = \frac{B_i}{B_k}$. By applying *Theorem 1* of [11] (see also [13] for more details), it can be shown that

$$R_k \sim \mathcal{N}(0, 1), \quad \text{Under } H_{0k}$$
$$R_k \sim \mathcal{N}(\mu_k, \tilde{\sigma}_{H_{1k}}^2), \quad \text{Under } H_{1k}$$

where $\mu_k = \sqrt{\frac{N_{dk} \beta_k}{\beta_k + 1}} \gamma_k$ and $\tilde{\sigma}_{H_{1k}} = 1 + \gamma_k$. The $P_f$ and $P_d$ of the test statistics (6) are thus given as [11]

$$P_{fk}(\lambda_k) = Pr\{R_k > \lambda_k | H_{0k}\} = \frac{1}{2} \text{erfc}\left(\frac{\lambda_k}{\sqrt{2}}\right) \quad (7)$$

$$P_{dk}(\lambda_k) = Pr\{R_k > \lambda_k | H_{1k}\} = \frac{1}{2} \text{erfc}\left(\frac{\lambda_k - \mu_k}{\sqrt{2} \tilde{\sigma}_{H_{1k}}}\right) \quad (8)$$

where $\lambda_k$ is the threshold and $\text{erfc}(.)$ is the complementary error function [15]. As we can see from (8), for the given $\gamma_k > 0$ and $\lambda_k$, increasing $N_{dk}$ increases $P_{dk}$. This is due to the fact that $\text{erfc}(.)$ is a decreasing function. Thus, the proposed detection algorithm is consistent and does not suffer from any SNR wall (i.e., for any given $P_{fk} > 0$ and $\gamma_k > 0$, $P_{dk} \to 1$ as $N_{dk} \to \infty$). One can also notice that the detector (6) is not very sensitive to small to medium interference signal. This is because, the ratio $\frac{\hat{M}_{dk}}{\hat{M}_z}$ will not be changed significantly in the presence of small to medium interference signals. Hence, the proposed detector is robust against small to medium ACI which will occur frequently in practice.

In the following, we address the relation between the detector (6) and the CED. As can be seen from (6), when $\beta_k \to \infty$, $\hat{M}_z$ becomes the true noise variance and the test statistics (6) will be

$$R_{k \beta_k \to \infty} = \sqrt{N_{dk}} \left( \frac{\hat{M}_{dk}}{\hat{M}_z} - 1 \right) = \sqrt{N_{dk}} \left( \frac{\hat{M}_{dk}}{\sigma^2} - 1 \right). \quad (9)$$

Indeed this is shifted and scaled version of the CED which is optimal. From this observation, we can understand that the role of $\hat{M}_z$ is just to estimate the noise variance from finite sub-band. Hence, the test statistics (6) can be considered as a GED.

Next we examine the following interesting question. For the given $\beta_k$, how much is the performance loss of (6)

compared to that of the CED? As can be seen from (7), $P_{fk}$ does not depend on $\beta_k$. Thus, the test statistics (6) and (9) will employ the same threshold $\lambda_k$ to ensure a certain $P_{fk}$. This threshold is given by

$$\lambda_k^\star = \sqrt{2}\text{erfc}^{-1}(2P_{fk}), \quad k \neq i.$$

Thus, the detection performance loss is given as

$$\eta_k = 1 - \frac{P_{dk}(R_k)}{P_{dk}(R_{k\beta_k \to \infty})} = 1 - \frac{\text{erfc}\left(\frac{\lambda_k^\star - \sqrt{\frac{N_{dk}\beta_k}{\beta_k+1}}\gamma_k}{\sqrt{2}(1+\gamma_k)}\right)}{\text{erfc}\left(\frac{\lambda_k^\star - \sqrt{N_{dk}}\gamma_k}{\sqrt{2}(1+\gamma_k)}\right)}.$$

From these explanations, the following key points can be highlighted:

1) If the noise variance is estimated from finite sub-band, the theoretical thresholds of the CED can not be applied directly.
2) When the bandwidth of the white sub-band is very small (i.e., $\beta_k$ is very small), the threshold value to ensure a certain $P_{fk}(P_{dk})$ of (6) is significantly different from that of the CED.
3) Increasing $\beta_k$ increases the detection performance of (6). Hence, the detection performance of the proposed GED is upper bounded by that of the CED.

## IV. SENSING TIME OPTIMIZATION

In this section, we compute the optimal $T_{ts}$ of (3) to maximize the throughput of the CR network (i.e., **Obj 2**). The CR network performs transmission in the $k$th sub-band when the GED (6) declares this sub-band as white. The proposed GED has a certain missed detection (i.e., the detector (6) may declare a non-white sub-band as white). Thus, in the $k$th sub-band, the CR network can have the following two SNRs [12]:

$$\gamma_c|H_{0k} = \gamma_c, \quad \text{Correct sensing decision}$$
$$\gamma_c|H_{1k} = \frac{\gamma_c}{1+\gamma_{pk}}, \quad \text{Incorrect sensing decision} \quad (10)$$

where $\gamma_c$ is the SNR of the CR network and $\gamma_{pk}, k \neq i$ are the SNR of the primary signal experienced at the receiver of the CR network[3]. If we denote the probability of the occurrences of $H_{0k}$ and $H_{1k}$ by $P(H_{0k})$ and $P(H_{1k})$[4], respectively, in the transmission time duration $T_f - T_o$ (recall Fig. 1), we will achieve the following two throughputs:

$$Th_{H_{0k}} = \frac{T_f - T_o}{T_f} R_{0k} P(H_{0k})(1 - P_{fk}(\tilde{\lambda}_k, T_o))$$
$$Th_{H_{1k}} = \frac{T_f - T_o}{T_f} R_{1k} P(H_{1k})(1 - P_{dk}(\tilde{\lambda}_k, T_o))$$

where $R_{0k} = \log_2(1+\gamma_c|H_{0k})$ and $R_{1k} = \log_2(1+\gamma_c|H_{1k})$[5]. Our objective will now be to get the optimal $T_o$ for maximizing the throughput of the $k$th

[3]In practice, we do not have any information about $\gamma_{pk}, k \neq i$. Due to this fact, we employ $\gamma_{pk} = \gamma_k, k \neq i$.
[4]These probabilities can be computed by employing the decision statistics of the previously sensed frames (i.e., the decision statistics of (6)).
[5]Here we assume that the CR network transmits a Gaussian signal and the channel between the CR transmitter and receiver is assumed to be AWGN.

sub-band under the constraint that the primary network (i.e., the $k$th sub-band) is sufficiently protected. This problem is mathematically formulated as

$$\max_{T_o} Th_{H_{0k}} + Th_{H_{1k}},$$
$$\text{s.t } P_{dk}(\tilde{\lambda}_k, T_o) \geq \tilde{P}_d, \quad k \neq i \quad (11)$$

where $P_{dk}(.)$ is the detection probability given in (8) and $\tilde{P}_d$ is the required detection probability. As can be seen from this expression, $P_{dk}$ depends on $\tilde{\lambda}_k$ and $T_o$. Furthermore, for the given $T_o$, the optimal $\tilde{\lambda}_k$ of the above problem can be obtained by setting [12]

$$P_{dk}(\tilde{\lambda}_k, T_o) = \tilde{P}_d, \quad k \neq i. \quad (12)$$

From (8), we will have

$$P_{dk}(\tilde{\lambda}_k) = \frac{1}{2}\text{erfc}\left(\frac{\tilde{\lambda}_k - \mu_k}{\sqrt{2}\tilde{\sigma}_{H_{1k}}}\right). \quad (13)$$

By combining (12) and (13), the optimal $\tilde{\lambda}_k$ becomes

$$\tilde{\lambda}_k^\star = a_k\sqrt{T_o} + b_k, \quad k \neq i \quad (14)$$

where $a_k = \sqrt{\frac{\beta_k B_k}{\beta_k+1}}\gamma_k$ and $b_k = \sqrt{2}\tilde{\sigma}_{H_{1k}}\text{erfc}^{-1}(2\tilde{P}_d)$. Substituting $\tilde{\lambda}_k^\star, k \neq i$ into (7) and after some straightforward steps, problem (11) can be reformulated as

$$\max_{T_o} \frac{T_f - T_o}{T_f}\left(\psi_k\text{erf}\left(\frac{a_k\sqrt{T_o} + b_k}{\sqrt{2}}\right) + \tilde{\psi}_k\right)$$
$$\triangleq \tilde{f}(T_o) \quad (15)$$

where $\psi_k = 0.5P(H_{0k})\log_2(1+\gamma_c|H_{0k})$ and $\tilde{\psi}_k = \psi_k + P(H_{1k})\log_2(1+\gamma_c|H_{1k})(1-\tilde{P}_d)$ are constants, and $\text{erf}(.) = 1 - \text{erfc}(.)$.

It can be shown that this problem is a concave maximization problem where its optimal $T_o$ can be obtained numerically by simple bisection search method (see Appendix B of [13] for more details).

Similar optimization problem has been considered in [12] by employing CED. However, the authors of [12] examine their problem by ignoring $Th_{H_1}$ (see equation (21) of [12]). The proposed generalized energy detection and sensing time optimization algorithms are summarized as follows.

**Algorithm I**: Generalized Energy Detection and Sensing Time Optimization
**Inputs**: $B$, $B_i$ and $B_k, k \neq i$ and $\gamma_k$.

1) **Obj 4**: Optimal sensing time computation
   a) From $B_i$ and $B_k, k \neq i$, determine the optimal $T_o^\star$ that maximizes the throughput by solving (15).
   b) Get the samples $T_o^\star B$ and denote the total received samples as $\{r[j]\}_{j=1}^{N_s}$ of (3), where $N_s = T_{ts}B$ with $T_{ts} = T_o^\star$.
2) **Obj 3**: Generalized energy detection
   a) Using these $N_s$ samples $\{r[j]\}_{j=1}^{N_s}$ and the white sub-band $B_i$, compute $R_k, k \neq i$ with (6).
   b) Compute $\lambda_k^\star$ using (14) and
      **if** $R_k < \lambda_k^\star$ **then**
         Label the $k$th sub-band as white.

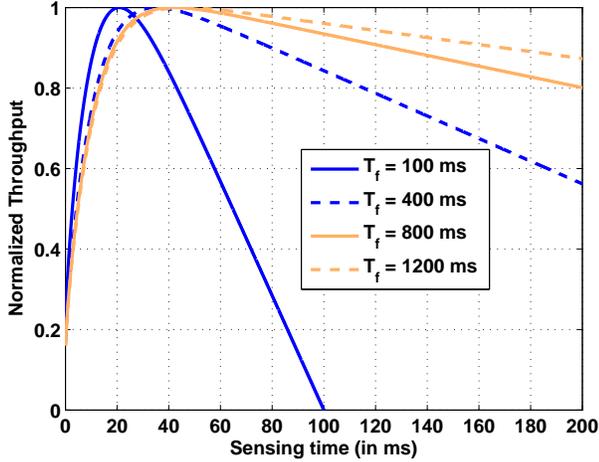

Fig. 2. Normalized $\tilde{f}(T_o)$ of (15) for different frame duration $T_f$

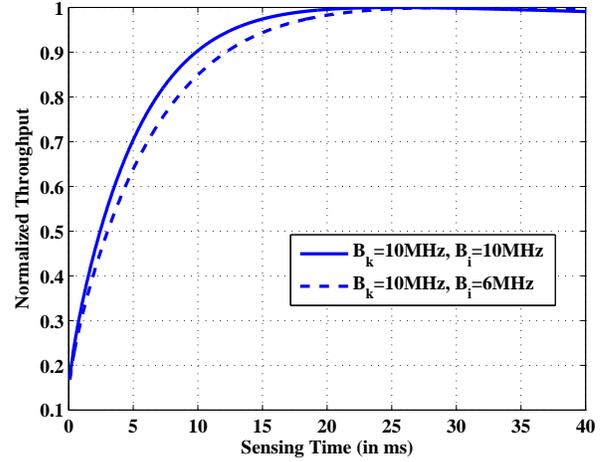

Fig. 3. Optimal sensing time for fixed $B_k$ and different $B_i$ and $T_f = 1.2$s.

**else**
    Label the $k$th sub-band as non-white.
**end if**

3) **Transmission**: If the $k$th sub-band is white, transmit information over the CR network in the remaining $T_f - T_o$ seconds.

Note that since we assume that $B_i$ is white, the CR can use both the $i$th and $k$th sub-band when the $k$th sub-band is white, otherwise the CR will use only the white sub-band $B_i$.

## V. NUMERICAL EXAMPLES

In this section we provide numerical examples on the sensing time optimization. Currently, the Federal Communications Commission (FCC) in USA has proposed TV bands for CR network application [12]. The bandwidth of each TV band is 6MHz and $\tilde{P}_d = 0.9$ at $\gamma_k = -20$dB. Suppose that we have a cognitive device with bandwidth 12MHz (i.e., 2 TV bands) where the first 6MHz is the target sub-band (i.e., sub-band $k$) and the remaining 6MHz is the white sub-band (i.e., sub-band $i$), $\gamma_c = 20$dB, $P(H_{0k}) = 0.8$ and $P(H_{1k}) = 0.2, k \neq i$. Under these settings, we plot the sensing time versus normalized throughput for different frame durations as shown in Fig. 2. As we can see when the frame duration increases, the optimal sensing time also increases. However, this increment is not linear. For example, the optimal sensing times with $T_f = 100$ms and $T_f = 1200$ms (i.e., 12 times increment) are 20ms and 45ms (i.e., 2.25 times increment), respectively. Thus, for practical application it is desirable to choose the maximum possible frame duration. For 802.22 system, we suggest to set $T_f = 2$ seconds.

In the following, we examine the effect of the bandwidths $B_k$ and $B_i$ on the optimal sensing time. To this end we take $B_k = B_i = 10$MHz, and $B_k = 10$MHz and $B_i = 6$MHz cases. Fig. 3 shows the sensing time versus normalized throughput for these two cases with frame duration $T_f = 1.2$s. From this figure, one can observe that the optimal sensing time with $B_k = B_i = 10$MHz setting (i.e., around 20ms) is less than that of the setting $B_k = 10$MHz, $B_i = 6$MHz (i.e., around 25ms). This figure also reveals the fact that increasing $B_i$ (i.e., $\beta_k$) will help to decrease the optimal sensing time. Also, from this figure and Fig. 2 we can notice that, for the given $B_i$, increasing $B_k$ will decrease the optimal sensing time which is expected.

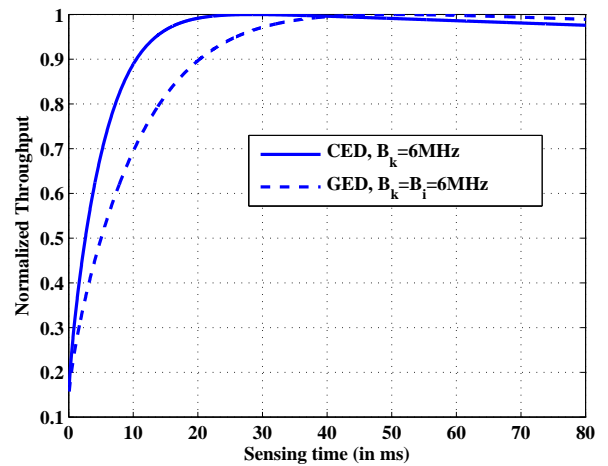

Fig. 4. Comparison of the sensing time of the CED and GED with $B_i = B_k = 6$MHz and frame duration $T_f = 2$s.

Next we compare the sensing time of the GED of (6) and that of the CED for the target sub-band with $T_f = 2$s which is shown in Fig. 4. As can be seen from this figure, the CED ($T_{oCED} = 28.5$ms) requires less sensing time compared to that of the proposed GED ($T_{oGED} = 50.6$ms) (i.e., $T_{oGED} \approx 1.78 T_{oCED}$). This is expected since the CED assumes perfect noise variance (i.e., $\beta_k \to \infty$). This result validates that the maximum throughput is achieved when the noise variance is known perfectly.

## VI. SIMULATION RESULTS

This section provides simulation results which are obtained by averaging 20000 experiments. Under $H_{1k}$ hypothesis, the signal samples are taken from the quadrature phase shift keying (QPSK) constellation with $\sigma_s^2 = 1$mw. Furthermore, in the $k$th sub-band (i.e., target sub-band), the channel between the primary transmitter and cognitive device is either AWGN or Rayleigh fading. The SNR is defined as $SNR \triangleq \frac{\sigma_s^2}{\sigma^2}$.

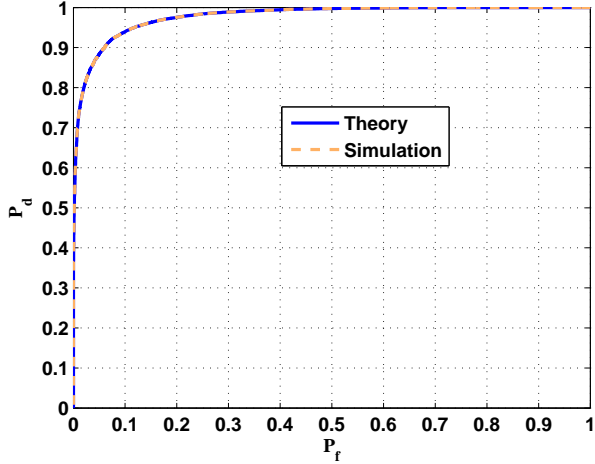 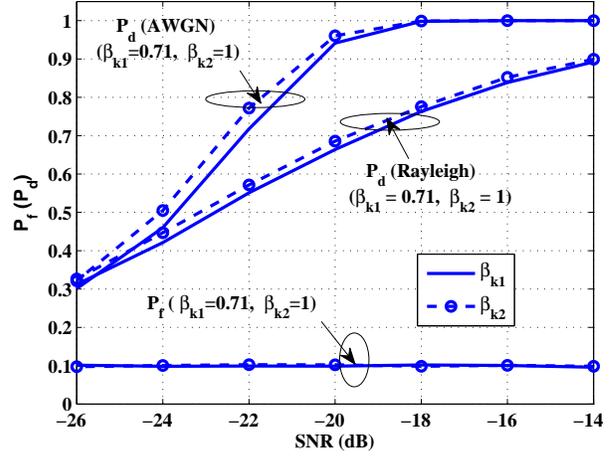

Fig. 5. Comparison of theoretical and simulated $P_f$ versus $P_d$ of GED in AWGN channel at SNR= $-20$dB.

Fig. 6. Performance of the proposed GED under noise variance uncertainty.

### A. Verification of $P_f$ versus $P_d$ Expressions of GED (6)

In this subsection, we verify the theoretical $P_{fk}$ and $P_{dk}$ expressions of the GED (6) by computer simulations. To this end, we set $B_i = 4.28$MHz, $B_k = 6$MHz (i.e., $\beta_k = 0.71$), $T_{ts} = 30.3$ms, SNR= $-20$dB and the channel between the primary transmitter and cognitive device is AWGN. Under these settings, Fig. 5 shows comparison of the theoretical and simulation $P_f(P_d)$. From this figure, we can see that the theoretical $P_f(P_d)$ matches exactly that of the simulated one.

### B. Effects of $\beta_k$ and Noise Variance Uncertainty on the $P_{fk}$ and $P_{dk}$ of GED

In this subsection, we examine the effects of $\beta_k$ and noise variance uncertainty on the $P_{fk}$ and $P_{dk}$ of GED. To this end, we consider that $B_i = [6, 4.28]$MHz, and $B_k = 6$MHz (i.e., $\beta_k = [\beta_{k1}, \beta_{k2}] = [1, 0.71]$). Furthermore, the true noise variance is modeled as a bounded interval of $[\frac{1}{\epsilon}\sigma^2 \ \epsilon\sigma^2]$ for some $\epsilon = 10^{\Delta\sigma^2/10} > 1$, where the uncertainty $\Delta\sigma^2$ is expressed in dB [4]. We assume that this bound follows a uniform distribution, i.e., $\mathcal{U}[\frac{1}{\epsilon}\sigma^2 \ \epsilon\sigma^2]$. The noise variance is the same for one experiment (since it has a short duration) and follows a uniform distribution during several experiments.

For this simulation, we set $T_{ts} = 30.3$ms, $\Delta\sigma^2 = 2$dB and $P_{fk1} = P_{fk2} = 0.1$, where $P_{fk1}$ and $P_{fk2}$ are the false alarm probabilities obtained by $\beta_{k1}$ and $\beta_{k2}$, respectively. Fig. 6 shows the achieved $P_f$ and $P_d$ for these $\beta_k$ values. From this figure, we can understand that the target $P_f \leq 1$ is maintained for both $\beta_k$ values. Thus, the $P_f$ of the detector (6) does not depend on the value of $\beta_k$ which is inline with the theoretical result. Furthermore, increasing $\beta_k$ (or SNR) increases the detection performance of the proposed GED (6) for both AWGN and Rayleigh fading channels. And, for a given $\beta_k$, the AWGN channel will achieve superior detection performance compared to that of the Rayleigh fading channel which is expected.

### C. Comparison of the Proposed GED and the detector of [11] for Pulse Shaped Signals

Recently new linear combination approach signal detection algorithm is proposed for pulse shaped transmitted signals with known rolloff factor in [11]. The detector of [11] is robust against noise variance uncertainty and small to medium ACI, and it outperforms CED and EVD-based signal detectors. Furthermore, the detector of [11] is already implemented using universal software radio peripheral (USRP) in [16] and has shown consistent result with the theory. Due to this reason, we compare the proposed GED (6) with the detector of [11] for pulse shaped signals with known rolloff factor (i.e., one band with known rolloff). To this end, we consider that the transmitted signal is pulse shaped by a square root raised cosine filter (SRRCF) with period $T_s$ and a certain rolloff factor (i.e., the total bandwidth of the transmitted signal is $\frac{(1+\text{rolloff})}{T_s}$Hz).

For the comparison, we consider the same scenario as in Fig. 3 of [11] (i.e., $T_s = \frac{1}{6}10^{-6}$s, rolloff $= 0.2$ and $T_{ts} = 4.55$ms). From fundamental wireless communication, it is known that the rolloff frequency regions of any pulse shaped signal are highly dominated by the noise (i.e., it contains almost noise only signal). Therefore, one can interpret that any pulse shaped transmitted signal has two sub-bands, where the first sub-band of bandwidth $\frac{1}{T_s}$Hz contains signal plus noise (under $H_1$ hypothesis) and the second sub-band of bandwidth $\frac{\text{rolloff}}{T_s}$Hz contains noise only signal. Hence, for our generalized energy detection algorithm (6), the former sub-band can be considered as the target sub-band (i.e., sub-band $B_k = 6$MHz$=[-3 : 3]$MHz) and the latter sub-band can be considered as the white sub-band (i.e., sub-band $B_i = 1.2$MHz$=[-3.6 : -3]$MHz and $[3 : 3.6]$MHz). Due to this reason, for the set up of Fig. 3 of [11], (6) utilizes $\beta_k = \frac{B_i}{B_k} = \text{rolloff} = 0.2$.

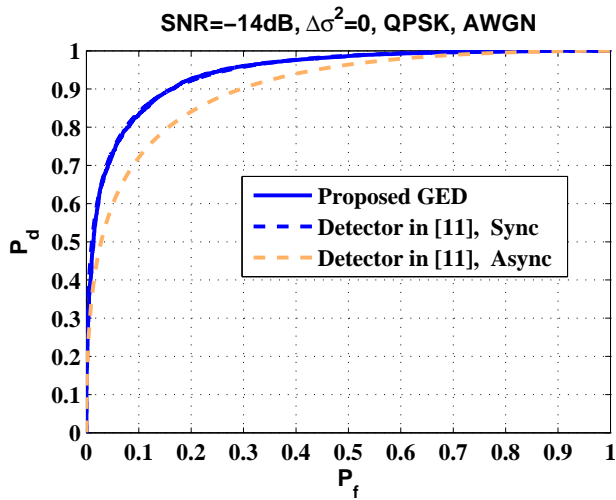

Fig. 7. Comparison of the proposed generalized energy detection algorithm and the detection algorithm of [11] for pulse shaped transmitted signals with $T_{ts} = 4.55$ms.

Fig. 7 shows the performances of the proposed GED and that of the detector in [11]. From this figure, we can see that the current algorithm achieves the same performance as that of [11] when the cognitive device is perfectly synchronized with the primary transmitter (i.e., Sync) which almost never happen in practice. However, for the practically relevant asynchronous scenario (i.e., Async), the current GED (which do not assume any synchronization) outperforms the detector in [11]. From this result, we can also conclude that the proposed GED outperforms CED and EVD based detectors under noise variance uncertainty. On the other hand, the proposed GED requires neither the pulse shaping filter of the transmitted signal nor oversampling of the received signal.

Like the detector in [11], the current GED does not depend on the phase of the received signal, and is not sensitive to carrier frequency offset and small to medium ACI signal. Thus, the proposed generalized energy detection algorithm is robust against frequency and phase offset, and small to medium ACI signal. The detailed analysis demonstrating this fact can be performed like that of [11].

## VII. Conclusions

This paper proposes novel spectrum sensing algorithm, and examines the sensing throughput tradeoff for CR networks under noise variance uncertainty. It is assumed that there are one white sub-band, and one target sub-band which is either white or non-white. Under this assumption, first we propose a novel GED for examining the target sub-band by exploiting the noise information of the white sub-band, then, we study the tradeoff between the sensing time and achievable throughput of the CR network. To study this tradeoff, we consider the sensing time optimization for maximizing the throughput of the CR network while ensuring that the primary network is sufficiently protected. The sensing time is optimized by utilizing the derived $P_d$ and $P_f$ expressions of the GED. The proposed GED does not suffer from SNR wall (i.e., robust against noise variance uncertainty). Numerical results reveal that the optimal sensing times with perfect and imperfect noise variances are not the same. Particularly, when the SNR= $-20$dB, frame duration 2s, and each of the bandwidths of the white and target sub-bands is 6MHz, we have found that the optimal sensing times are 28.5ms and 50.6ms with perfect and imperfect noise variance scenarios, respectively. The derived $P_d$ and $P_f$ expressions of the GED are verified by computer simulation. Simulation results also demonstrate that the proposed GED is robust against noise variance uncertainty and outperforms the existing signal detectors.